\documentclass[twocolumn,pra,aps,showpacs]{revtex4}

\usepackage{amsmath}
\usepackage{amssymb}
\usepackage[dvips]{graphicx}
\usepackage{pstricks}
\usepackage{pst-node}
\usepackage{pst-plot}

\newcommand{\qed}{\hspace*{\fill}$\square$}

 \newtheorem{thm}{Theorem}

 \newtheorem{defn}[thm]{Definition}
 \newtheorem{prop}[thm]{Proposition}
 \newenvironment{proof}{\noindent \emph{Proof.}}{\qed}

 \newcommand{\Z}{\mathbf{Z}}
 
 \newcommand{\Zd}{\Z_2}
 \newcommand{\Zdn}{\Zd^n}
 \newcommand{\Zddn}{\Zd^{2n}}

 \newcommand{\paratodo}{\forall\,}
 
 \newcommand{\vect}[1]{\boldsymbol{\mathrm{#1}}}
 \newcommand{\set}[2]{ \{\,#1\,|\,#2\,\}}
 \newcommand{\sset}[1]{ \{#1\} }
 
 \newcommand{\transp}{^t}

 \newcommand{\sumaconexa}{\,\sharp\,}

 \newcommand{\fase}[1]{\varphi(#1)}

 \newcommand{\ket}[1]{|#1\rangle}

 \newcommand{\Dn}{{\cal H}_2^{\otimes n}}
 \newcommand{\err}{\mathcal E}
 \newcommand{\Pauli}[2]{\mathbf P_{#1}(#2)}
 \newcommand{\paulis}{\sigma}
 \newcommand{\pauli}[1]{\paulis_{#1}}
 \newcommand{\pauliv}[1]{\pauli {\vect #1}}
 \newcommand{\paulid}[2]{\pauli {#1 #2}}
 \newcommand{\paulidv}[2]{\paulid {\vect {#1}} {\vect {#2}}}

\begin{document}

\title[Short Title]{
Topological Quantum Error Correction with Optimal Encoding Rate}

\author{H. Bombin and M.A. Martin-Delgado}
\affiliation{
Departamento de F\'{\i}sica Te\'orica I, Universidad Complutense,
28040. Madrid, Spain.
}

\begin{abstract}
We prove the existence of topological quantum error correcting
codes with encoding rates $k/n$ asymptotically approaching the
maximum possible value. Explicit constructions of these
topological codes are presented using surfaces of arbitrary genus.
We find a class of regular toric codes that are optimal. For
physical implementations, we present planar topological codes.

\end{abstract}

\pacs{03.67.-a, 03.67.Lx}

\maketitle

\begin{figure}
\psset{xunit=7cm,yunit=3cm}
\begin{pspicture}(-.3,-.2)(0.82,1)
\rput(-.1,.45){$k/n$} %
\rput(.27,-.2){$t/n$} %
% ejes
\psset{xunit=5\psxunit}%
%labels
\psset{Dy=1, Dx=.1}%%
\psaxes[axesstyle=none,ticks=none,labelsep=1.4\pslabelsep](0,0)(.11,1)%
% box
\psset{Dy=.2, Dx=.02}%
\psaxes[axesstyle=frame,tickstyle=top, labels=none](0,0)(.11,1)
% Hamming
\psplot[plotpoints=100,linewidth=1.5\pslinewidth]{0.000001}{.11}{1
x 2 log 3 log div mul x
log 2 log div x mul neg 1 x sub 1 x sub log 2 log div mul sub add sub}%
\psset{plotstyle=dots, dotscale=1.5}%%tamaño de las marcas
% codigo origen
\psdot[dotstyle=o](.1,.2)
% codigos toricos opt%
\parametricplot[dotstyle=triangle,plotpoints=20]
{21}{2}{t 2 t mul 1 add 2 t mul 1 add mul 1 add div 2 2 t mul 1
add 2 t mul 1 add mul 1 add div}
% codigos K_t t = 4l+1
\parametricplot[dotstyle=square,plotpoints=20]%
{85}{9}{1 t 1 sub t mul 2 div div t 1 sub t mul 2 div t 1 sub 2
mul sub t 1 sub t mul 2 div div}
\end{pspicture}
\caption{The rate $k/n$ vs. $t/n$ for the optimized toric codes
($\triangle$) and for the codes derived from self-dual embeddings
of graphs $K_{4l+1}$ ($\square$); $\bigcirc$ corresponds to the
embedding of $K_5$ in the torus. The quantum Hamming bound is
displayed as a reference (solid
line).}\label{figura_rates_cuanticos}
\end{figure}
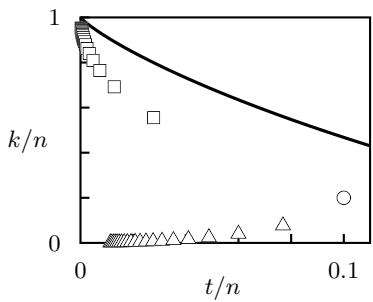

\section{Introduction}

Quantum computation has overcome major difficulties and has become
a field of solid research. On the theoretical side, several models
of quantum computation are already proposed like the quantum network
model using a set of universal logic gates. Quantum error correction
and fault tolerant quantum computation have been proved to be well
established theoretically. On the experimental side, test-ground
experiments have been conducted with a small number of quantum logic gates
based on several proposals for realizing qubits. These constitute
proof-of-principle experimental realizations
showing that theory meets experiment.

Yet, it still faces a major challenge in order to built a real
quantum computer: for a scalable quantum computer to be ever built,
we have to battle decoherence and systematic errors in an efficient way
\cite{shor1}, \cite{steane}.
In fact, the network model corrects errors combinatorially and
this requires a very low initial error rate, known as the threshold,
in order to stabilize a quantum computation
\cite{shor2}, \cite{knill_etal}, \cite{kitaev1}, \cite{aharonov}.

There exists a very clever proposal of fault-tolerant quantum computation
based on quantum topological ideas \cite{kitaev2}, \cite{freedman_etal}.
The idea is to design the
quantum operations so as to have a physically built-in
mechanism for error correction, without resorting to
external corrections every time an error occurs \cite{dennis}.
The key point here is that quantum topology is a global resource
that is robust against local errors, thereby providing a natural setup
for fault tolerance \cite{preskill}.

\section{Topological Codes on Arbitrary Surfaces}

A prerequisite for a topological QC is a topological quantum code
for error detection and correction. It serves also as a
quantum memory.
In addition, quantum error correction codes are useful for
quantum communication channels
while sharing the feature of being quantumly robust.

A Hamiltonian can be constructed such that its ground state coincides
with the code space.
The nice thing of topological quantum codes (TQC) is that the generators
are local and this makes feasible the experimental implementation of these
codes, although other obstacles have to be overcome as we shall explain.

In this paper we shall provide explicit constructions of topological
quantum codes with encoding rates beating those that can be
achieved with current toric codes \cite{kitaev2}
(see Fig.~\ref{figura_rates_cuanticos} and eq. \eqref{limit}).

A \emph{quantum error correcting code of length $n$} is a subspace
$\mathcal C$ of $\Dn$, with ${\cal H}_2$ the Hilbert space of one
qubit, such that recovery is possible after noise consisting of
any combination of error operators from some set $\err$ of
operators on $\Dn$. The set $\err$ is the set of \emph{correctable
errors}, and we say that $\mathcal C$ \emph{corrects} $\err$. For
codes of length $n$, let $\err (n,k)$ be the set of operators
acting on at most $k$ qubits. We define the \emph{distance} of the
code $\mathcal C$, denoted $d(\mathcal C)$, as the smallest number
$d$ for which the code does not detect $\err(n,d)$. A code
$\mathcal C$ corrects $\err(n,t)$ iff $d(\mathcal C)>2t$. In this
case we say that $\mathcal C$ corrects $t$ errors. We talk about
$[[n,k,d]$] codes when referring to quantum codes of length $n$,
dimension $2^k$ and distance $d$. Such a code is said to encode
$k$ qubits. The {\em encoding rate} is $\frac{k}{n}$.

We consider the following family of operators acting on a string
of qubits of length $n$:
\begin{equation}
 \pauliv v := \paulidv x z := \bigotimes_{j=1}^n i^{x_j z_j} X^{x_j}Z^{z_j},
\end{equation}
where $\vect x,\vect z\in\Zdn$, $\vect v = (\vect x \vect z):=
(x_1,\dots,x_n,z_1,\dots,z_n)$ and $X,Z$ are the standard Pauli
matrices. They commute as follows:
\begin{equation}\label{ConmutacionSigmas}
\pauliv u\pauliv v=\fase{\vect u\transp\Omega \vect v}\pauliv v \pauliv u %
\end{equation}
where \begin{equation}\Omega:=
\begin{bmatrix}
0 & 1 \\ -1 & 0
\end{bmatrix}\end{equation}
is a $2n\times 2n$ matrix over $\Zd$ and
$\fase{k} := {\rm e}^{\pi {\rm i}k}$, $k\in\Zd$.
We keep the minus sign in $\Omega$ because it appears for higher
dimensionality or qudits \cite{numbertheory}.
 The group of all the operators generated
by the set of $\paulis$-operators is the Pauli group $\Pauli D n$.

There exists a nice construction using the Pauli group to
construct the  symplectic or stabilizer codes
\cite{gottesman}, \cite{calderbank_etal}.
For any subspace $V\subset\Zddn$, we define the subspace
$\widehat{V}:=\set{\vect u \in\Zddn}{\paratodo\vect v\in
V\,\,\vect v\transp\Omega \vect u = 0}$. %\end{equation}
Let $V_{\mathcal C}\subset\Zddn$ be any \emph{isotropic} subspace,
that is, one verifying $V_{\mathcal C}\subset \widehat V_{\mathcal
C}$ \cite{numbertheory}. Let $B$ be one of its basis and set
${\cal S}=\set{\pauliv v}{\vect v \in B}$. The code
\begin{equation}
    \mathcal C := \set{\ket \xi \in \Dn}{\paratodo \paulis \in {\cal S}\,\,
\paulis \ket \xi = \ket \xi }
\end{equation}
detects the error $\pauliv u$ iff $\vect u\;\not\in\; \widehat
V_{\mathcal C}-V_{\mathcal C}$ and thus its distance is
\begin{equation}d(\mathcal C)= \min_{\vect u\in{\widehat V_{\mathcal C}-V_{\mathcal C}}} |\vect u|,
\label{distance}
\end{equation}
where the norm is just the number of qubits on which $\pauliv u$
acts nontrivially. The set ${\cal S}$ is the stabilizer of the
code. This way, the problem of finding good codes is reduced to
the problem of finding good isotropic subspaces $V_{\mathcal C}$.

\begin{figure}
 \includegraphics[width=8.5cm]{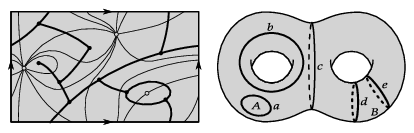}
 \caption{\emph{Left.} A graph (thick lines) in the torus and its dual
(weak lines). \emph{Right.} Several cycles in the 2-torus. $a$ is
the boundary of $A$, and $c$
 is also homologous to zero because it encloses half of the surface. $b,d,e$ are not
 homologous to zero. $d$ and $e$ are homologous because they enclose the area $B$. }\label{figura_duales}
\end{figure}

Thus far we have dealt with the purely algebraic structure of
codes. Now, we turn to the connection with topology.
 For a \emph{surface} we understand a compact connected
2-dimensional manifold. Well known examples of surfaces are the
sphere $S$, the torus $T$ and the projective plane $P$. More
generally one can consider the $g$-torus or sphere with $g$
handles $gT$, and the sphere with $g$ crosscaps $gP$. $gT$ and
$gP$ are said to have \emph{genus} $g$. In fact, it is a well
known result of surface topology that the previous list of
surfaces is complete.

We shall not be restricted to codes based on regular lattices on
a torus, or toric codes \cite{kitaev2}, but we shall use general
graphs embedded in surfaces of arbitrary genus
in order to explore optimal values
of the encoding rate $\frac{k}{n}$.
A \emph{graph} $\Gamma$ is a collection of \emph{vertices} $V$ and
 \emph{edges} $E$. Each edge joins two vertices. A graph is usually visualized
flattened on the plane.
%as a set of points (vertices) in the plane  plus a collection of
%arcs joining them (edges).
Now consider a graph $\Gamma$ embedded in a surface $M$ (see
Fig.~\ref{figura_duales}). When $M-\Gamma$ is a disjoint
collection of discs, we say that the embedding is a {\em cell
embedding}. Let us gather these discs into a set of \emph{faces}
$F$.

 We now introduce the notion of a \emph{dual graph}, which
is crucial for the construction of quantum topological codes.
Given a cell embedding $\Gamma_M$ of a graph $\Gamma$ in a surface
$M$, the dual embedding $\Gamma^\ast_M$ is constructed as follows.
For each face $f$ a point $f^\ast$ is chosen to serve as a vertex
for the new graph $\Gamma^\ast$. For each edge $e$ lying in the
boundary of the faces $f_1$ and $f_2$, the edge $e^\ast$ connects
$f_1^\ast$ and $f_2^\ast$ and crosses $e$. Each vertex $v$
corresponds to a face $v*$. The idea is illustrated in
Fig.~\ref{figura_duales}.

Let us enlarge a bit the concept of surface. Take a surface $M$
and delete the interiors of a finite collection of non-overlapping
discs. We say that the resulting space is a \emph{surface with
boundary}. Note that for any cell embedding in such a surface, the
boundary is a subset of the embedded graph. One could argue that
no dual graph can be defined for this surfaces, but in fact this
is not a major difficulty. It is enough to think that some of the
vertices in the dual graph are 'erased'. For us the most important
example of surface with boundary will be the $h$-holed disc $D_h$,
$h\geq 1$.

Consider a cell embedding $\Gamma_M$ on a surface, with or without
boundary. In order to construct the physical system realizing the
topological code $\mathcal C$, we attach a qubit to each edge of
the graph $\Gamma$. The study of the stabilizer and correctable
errors of the code gets benefited by using $Z_2$ homology theory,
which we shall now introduce. Consider a $Z$-type operator
$\paulidv 0 \lambda$. A chain is a formal sum
 of edges $c_1=\sum_j \lambda_j\, e_j$. We relate chains and $Z$-type
 operators setting $\pauli {c_1}:=\paulidv 0 \lambda$. Similarly, given a
 cochain or formal sum of dual edges $c^1=\sum_j \lambda'_j\,e_j^\ast$,
 we relate it to an $X$-type operator setting  $\pauli
{c^1}:=\paulidv {\lambda'} 0$. There is a natural product between chains and cochains,
 namely $(c^1,c_1) := \vect \lambda \cdot \vect \lambda'$.
The point is that
\begin{equation}\label{propiedad_conmutador_vs_producto}
\pauli {c^1}\pauli {c_1}=\fase {(c^1,c_1)}\,\pauli {c_1}\pauli
{c^1}.
\end{equation}
This expression already shows that the commutation relations of
operators are determined by the topology of the cell embedding
$\Gamma_M$.
 Compare with \eqref{ConmutacionSigmas}.

Note that a chain is nothing but a collection of edges, those with
a coefficient equal to one, and thus can be easily visualized as
lines drawn on the surface. If a chain has an even number of edges
at every vertex, we call it a \emph{cycle}. When a cycle encloses
an area of the surface, we say that it is a \emph{boundary}. If
two cycles enclose an area altogether, they are said to be
\emph{homologous}. Boundaries are homologous to zero. Figures
\ref{figura_duales} illustrate these concepts. \emph{Cocycles} are
defined analogously but in the dual graph. \emph{Coboundaries} are
a bit different, however, at least in the case of surface with
boundary. If cutting the surface along a cocycle divides it apart
into two pieces, then it is a coboundary. Given a face $f$, we
shall denote by $\partial f$ its boundary. Similarly, given a dual
face $v^\ast$ we denote its coboundary by $\delta v^\ast$, see
Fig.~\ref{figura_homologia_disco}.

Before stating the main result about general constructions of
topological quantum codes for arbitrary graphs embedded in
surfaces, we need a pair of ingredients. Given a a surface $M$
there exists always some cell embedding on it. The \emph{Euler
characteristic} of $M$ is defined by
\begin{equation}
\chi(M):=|V|-|E|+|F|,
\end{equation}
and it does not depend upon the embedded graph. Now, let
$\Gamma_M$ be a cell embedding of a graph $\Gamma$ in a surface
$M$. We define the distance $\mathrm d(\Gamma_M)$ as the minimal
length (edge amount) among those cycles which are not homologous
to zero. For surfaces with boundary, $\mathrm d(\Gamma^\ast_M)$
should be understood as a symbol denoting the minimal length among
cocycles.

\begin{thm} \textbf{Topological codes. }
 Let $\Gamma_M$ be a cell embedding of a graph in a surface.
 The symplectic code $\mathcal C$  of length $n=|E|$ with stabilizer
${\cal S}=\set{\pauli{\delta v^\ast}}{v\in V}\cup
 \set{\pauli{\partial f}}{f\in F}$ has distance
$d=\min\sset{\mathrm d(\Gamma_M), \mathrm d(\Gamma^\ast_M)}$ and
 encodes $k=2-\chi(M)$ qubits if $M$ does not have any boundary or
$k=1-\chi(M)$ qubits if it does.
 \end{thm}

\begin{proof} This proof involves homology theory. Following standard notation,
 we denote the first homology group by $H_1=Z_1/B_1$ and the first cohomology group
 by $H^1=Z^1/B^1$. Now, since $(\delta
v^\ast,\partial f)=0$, the space $V_{\mathcal C}$ is isotropic.
Note that $V_{\mathcal C}\simeq B^1\oplus B_1$. A key observation
is that $(\delta v^\ast,c_1)=0$ iff $c_1\in Z_1$, and similarly
for $\partial f$ and $Z^1$. In other words, $\hat V_{\mathcal
C}\simeq Z^1\oplus Z_1$, and the distance \eqref{distance} is the
one stated. As $\dim \hat V_{\mathcal C} - \dim V_{\mathcal C} =
2k$, where $k$ is the number of encoded qubits, it only remains to
know the dimension of the homology and cohomology group. But we
have $H_1\simeq H^1\simeq\Zd^{2-\chi}$ for surfaces without
boundary and $H_1\simeq H^1\simeq\Zd^{1-\chi}$ for surfaces with
boundary.
\end{proof}

Since $\chi(gT)=2-2g$, the $g$-torus yields codes with $k=2g$
logical qubits. $\chi(gP)=2-g$, and thus codes from $gP$ encode
$k=g$ qubits. For the $h$-holed disc $D_h$, $\chi(D_h)=1-h$ and
$k=h$. The parity check matrix $H$ of a topological code has the
diagonal form $diag (H_1,H_2)$ where the matrices $H_1$ and $H_2$
are in essence the incidence matrices of $\Gamma$ and
$\Gamma^\ast$. Thus, topological codes are an example of
generalized CSS codes \cite{css1}, \cite{css2}.

Note that uncorrectable errors are related to cycles which are not
homologous to zero. This is exemplified as part of
Fig.~\ref{figura_homologia_disco}. Therefore, the whole problem of
constructing good topological codes related to a certain surface
relies on finding embeddings of graphs in such a way that both the
embedded graph \emph{and} its dual have a big distance whereas the
number of edges keeps as small as possible. Thus, we find that
this quantum problem can be mapped onto a problem of what is
called {\em extremal topological graph theory}, a branch devoted
to graph embeddings on surfaces \cite{tgt} and the computation of
maxima/minima of certain graph properties. To this end, we find
very useful to introduce the following concept.

\begin{defn} Given a surface $M$ and a positive
integer $d$, we let the quantity $\mu(M,d)$ be the minimum number
of edges among the embeddings of graphs in $M$ giving a code of
distance $d$. \end{defn}

Since calculating the value of the function $\mu$ is a hard
problem, we shall investigate some of its properties. The issue
of connectivity between sites suggests also the introduction of a
refinement of $\mu$. The quantity $\mu_c(M,d)$ is defined as
$\mu(M,d)$ but with the restriction that the graphs can have faces
with at most $c$ edges and vertices lying in at most $c$ edges. By
connectivity here we mean that vertex $\pauli{\delta v^{\ast}}$
and face $\pauli{\partial f}$ operators should act over at most
$c$ qubits. A loosely connected system simplifies the error
correction stage.

\section{Optimal Encoding Rates}

\begin{figure}\includegraphics[width=7cm]{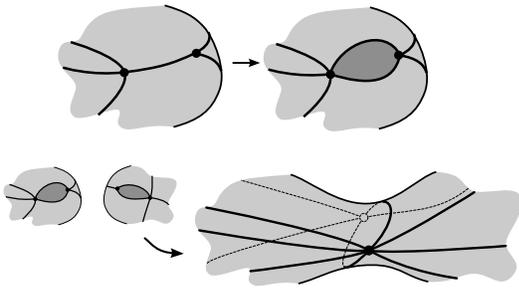}
 \caption{The construction that proves the topological subadditivity of $\mu$.
 The first step is to perform a cut along a selected edge in each of the embeddings to be connected.
 Then the resulting boundaries must be identified.}
 \label{figura_suma_codigos}
\end{figure}

There is an interesting result in surface topology stating that
every surface can be obtained by combination of $S$, $P$ and $T$.
To perform this connection over two surfaces, say $M_1$ and $M_2$,
one chooses two discs $D_i\subset M_i$. The \emph{connected sum}
of $M_1$ and $M_2$, denoted $M_1\sumaconexa M_2$, is constructed
by deleting the interiors of $D_1$ and $D_2$ and identifying its
boundaries. Connecting $g\geq 0$ tori to a sphere one gets $gT$,
and connecting $g\geq 1$ projective planes one gets $gP$. The
point is that given embeddings of distance $d$ in $M_1$ and $M_2$,
a new embedding can be constructed in $M_1\sharp M_2$ in such a
way that the number of edges does not increase and the distance is
preserved. The procedure is displayed in
Fig.~\ref{figura_suma_codigos}. This implies the following result,
which we shall use to proof asymtoptic properties about minimal
sizes of topological codes:

\begin{prop}
\textbf{Topological Subadditivity.} Given two surfaces $M_1$ and
$M_2$
\begin{equation}
\mu(M_1\sharp M_2,d)\leq \mu(M_1,d)+\mu(M_2,d).
\end{equation}
\end{prop}

\begin{figure}\includegraphics[width=8.5cm]{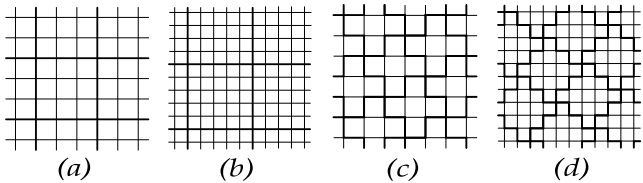}
 \caption{Four self-dual lattices on the torus. Here the torus is represented as a quotient
  of the plane and a tessellation of it.
  Thick lines are the border of tesserae.
  \emph{(a),(b):} The toric codes introduced in \cite{kitaev2}, for $d=3$ and $d=5$.
  \emph{(c),(d):} The optimal regular toric codes for $d=3$ and $d=5$.}
 \label{figura_toric_codes}
\end{figure}

Let us apply these tools starting with the torus $T$, the simplest
orientable surface with nontrivial first homology group. In
\cite{kitaev2}, a family of toric codes was presented, in the form
of self-dual regular lattices on the torus. This is a very simple
instance of topological graph theory. One can consider other
self-dual regular lattices embedded on the torus. All of them
share the property that vertex $\pauli{\delta v^{\ast}}$ and face
$\pauli{\partial f}$ operators act on $c=4$ qubits. Among them, we
have found an optimal family of lattices that demand half the
number of qubits. Examples of both systems of lattices are
depicted in Fig.~\ref{figura_toric_codes}, were the torus is
represented as a quotient of the plane through a tessellation. The
original toric codes lead to a family of $[[2d^2,2,d]]$ codes
\cite{kitaev2}. Our lattices give $[[d^2+1,2,d]]$ codes. This
already shows that
\begin{equation}\mu(T,d)\leq\mu_4(T,d)\leq d^2+1.\end{equation}
Invoking topological subadditivity, we learn that $\mu(nT,d)$ is $O(d^2)$
%in its second argument
, that is, it grows at most quadratically with
$d$.

This analysis shows that our toric lattices yield a better encoding rate
$\frac{k}{n}\sim \frac{2}{d^2}$ than the original ones with
$\frac{k}{n}\sim \frac{1}{d^2}$.
However, despite being optimal, these toric lattices
produce encoding rates  vanishing
in the limit of large $n$ qubits (see Fig.~\ref{figura_rates_cuanticos}).
Then, a major challenge arises:
is it possible to find topological quantum codes with non-vanishing encoding
rates? And what about  approaching the maximum value of 1?
The answer is positive in both cases and we hereby show the construction.

\begin{figure}
\includegraphics[width=8.5cm]{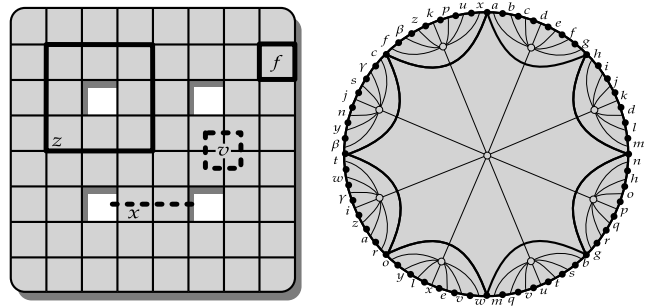}
 \caption{\emph{Left.} A graph embedding in the surface $D_4$ yielding a
code of distance 3.
 We display a boundary $\partial f$, a coboundary $\delta v^\ast$ and two uncorrectable
 errors of $Z$ and $X$-type: the  cycle $z$ is not a boundary, and the cocycle $x$ is not a coboundary.
 \emph{Right.} The self-dual embedding of $K_9$ in the 10-torus. Thick lines represent the graph, weak ones its dual.}
 \label{figura_homologia_disco}
\end{figure}

To this end, let us introduce the
complete graph $K_s$ as the graph consisting of $s$ vertices
and all the possible edges among them.
 A closer examination of
Fig.~\ref{figura_toric_codes} reveals that the lattice giving a
$[[10,2,3]]$ code is a self-dual embedding of $K_5$. This suggests
considering self-dual embeddings of $K_s$, since such an embedding
would give a $[[\binom s 2, \binom s 2 -2(s-1), 3]]$ code. In
fact, these embeddings are possible in orientable surfaces with
the suitable genus as long as $s=1 \ (\text{mod } 4)$ \cite{tgt}.
In Fig.\ref{figura_duales} we show the constructions of such an
embedding. Due to topological subadditivity and the fact that
$\mu(gT,d)\geq\mu(gT,1) = 2g$, the family of codes given by
self-dual embeddings of complete graphs $K_s$ is enough to show
that for $d=3$
\begin{equation}
\lim_{n\rightarrow \infty} \frac{k}{n} = \lim_{g\rightarrow
\infty} \frac {2g}  {\mu(gT,3)} =1,
\label{limit}
\end{equation}
that is, the ratio $k/n$ is asymptotically one, and thus good
topological codes can be constructed, at least in the case of
codes correcting a single error. Figure
\ref{figura_rates_cuanticos} displays the rates for this family of
codes and also for the optimized toric codes. Now we can
appreciate the very different behaviour between toric codes and
topological codes embedded in higher genus surfaces, the latter
ones allowing us to increase the encoding rate up to its maximal
value.

So far we have not touched upon the question of physical
implementations. Consider the system of qubits arranged according
to a given graph embedding, as explained above. The hamiltonian
\begin{equation}
H = -\sum_{f\in F} \paulis_{\partial f}-\sum_{v\in V}
\paulis_{\delta v^\ast}
\end{equation}
has a degenerate ground state whose elements are the protected
codewords. This system is naturally protected against
errors \cite{kitaev2}. A major drawback is the requirement that the
physical disposition of the qubits must give rise to a nontrivial
topology. It is difficult to imagine an experimentalist
constructing a system living in a torus, for example.

However, the formalism that we have presented allows us to use
surfaces with boundary. In particular, the $h$-holed disc $D_h$
gives rise to codes encoding $k=h$ qubits but has the advantage of
being a subset of the plane. Figure \ref{figura_homologia_disco}
shows an example of a very regular embedding in
$D_4$. It is apparent how to generalize this example to a higher
number of encoded qubits $k$ and distances $d$. As in the case of surfaces
without boundaries, the length of the code will scale as $O(k
d^2)$.

\section{Conclusions}

Finally, we want to emphasize that the locality of topological
codes  is a very important issue for their implementation in
physical systems, now more feasible after the introduction of
planar topological codes. The embeddings of complete graphs
provide encoding rates that overcome the barrier established so
far by toric codes. Moreover, we have introduced a measure $\mu $
that establishes an interplay between quantum information and
extremal topological graph theory.

\noindent {\em Acknowledgements}
We acknowledge financial support from a
PFI fellowship of the EJ-GV (H.B.),
DGS grant  under contract BFM 2003-05316-C02-01 (M.A.MD.),
and CAM-UCM grant under ref. 910758.

\end{document}